\documentclass[fleqn,usenatbib,times]{mnras}

\usepackage{newtxtext,newtxmath}
\usepackage{graphicx}
\usepackage{tikz}
\usepackage{wasysym}
\usepackage{hyperref}
\def\equationautorefname~#1\null{(#1)\null}
\usepackage[T1]{fontenc}
\usepackage{ae,aecompl}

\usepackage{cleveref}
\crefformat{section}{\S#2#1#3} 
\crefformat{subsection}{\S#2#1#3}
\crefformat{subsubsection}{\S#2#1#3}

\usepackage{graphicx}	
\usepackage{amsmath}	
\usepackage{amssymb}	

\title[Pre-impact spin in Moon-forming collisions]{The effect of pre-impact spin on the Moon-forming collision}

\author[S. Ruiz-Bonilla et al.]{
S. Ruiz-Bonilla$^{1,2}$\thanks{E-mail: sergio.ruiz-bonilla@durham.ac.uk},
V. R. Eke$^{1},$
J. A. Kegerreis$^{1},$
R. J. Massey$^{1},$
L. F. A. Teodoro$^{3,4}$
\\
$^{1}$Institute for Computational Cosmology, Durham University, South Road, Durham DH1 3LE, UK\\
$^{2}$Institute for Data Science, Durham University, South Road, Durham DH1 3LE, UK\\
$^{3}$BAERI/NASA Ames Research Center, Moffett Field, CA, USA\\
$^{4}$School of Physics and Astronomy, University of Glasgow, G12 8QQ, Scotland, UK
}

\date{Accepted XXX. Received YYY; in original form ZZZ}

\pubyear{2020}

\begin{document}
\label{firstpage}
\pagerange{\pageref{firstpage}--\pageref{lastpage}}
\maketitle

\begin{abstract}
We simulate the hypothesised collision between the proto-Earth and a Mars-sized impactor that created the Moon. Amongst the resulting debris disk in some impacts, we find a self-gravitating clump of material. It is roughly the mass of the Moon, contains $\sim1\%$ iron like the Moon, and has its internal composition resolved for the first time. The clump contains mainly impactor material near its core but becomes increasingly enriched in proto-Earth material near its surface. A graduated composition has recently been measured in the oxygen isotope ratios of Apollo samples, suggesting incomplete mixing between proto-Earth and impactor material that formed the Moon. However, the formation of the Moon-sized clump depends sensitively on the spin of the impactor. To explore this, we develop a fast method to construct models of multi-layered, rotating bodies and their conversion into initial conditions for smoothed particle hydrodynamical (SPH) simulations. We use our publicly available code to calculate density and pressure profiles in hydrostatic equilibrium, then generate configurations of over a billion particles with SPH densities within 1\% of the desired values. This algorithm runs in a few minutes on a desktop computer, for $10^7$ particles, and allows direct control over the properties of the spinning body. In comparison, relaxation or spin-up techniques that take hours on a supercomputer before the structure of the rotating body is even known. Collisions that differ only in the impactor's initial spin reveal a wide variety of outcomes: a merger, a grazing hit-and-run, or the creation of an orbiting proto-Moon.

\end{abstract}




\section{Introduction}
\label{introduction}

From planets and stars to dark matter haloes, self-gravitating spinning objects are common in astronomy. Their spin reflects the particular history of gravitational torques experienced by the material that they contain. By studying the angular momenta of astronomical systems, we can learn about the processes through which these objects formed. 

As self-gravitating objects can only spin so fast without breaking apart, the orbital angular momentum of accreting material typically dominates over that present due to spin. For instance, the final stage of planet formation involves giant impacts between planet-sized bodies \citep{Chambers+Wetherill1998, Clement+2019}, and the pre-impact spins are usually ignored despite the fact that rapidly rotating bodies are a common outcome of such collisions \citep{Kokubo+2010,Li+2020}. Examples of this include attempts to explain Uranus' rotation axis using orbital angular momentum brought by a $2$--$3\, M_\oplus$ object \citep{Slattery+1992, Kegerreis+2018}, and models in which the angular momentum of the Earth--Moon system results from the impact of a non-rotating Mars-sized body, Theia, and a non-rotating proto-Earth \citep{Canup+2001}. 

The Moon-forming impact is one planetary example for which pre-impact spin has received consideration. \citet{Canup2008} showed how pre-impact rotation changed the collision outcome relative to the canonical impact studied by \citet{Canup+2001}. The isotopic similarity of the Earth's mantle and lunar samples \citep{Wiechert+2001} provoked attempts to place a higher fraction of proto-Earth material into the protolunar disc by starting with a spinning target \citep{Cuk+Stewart2012, Lock+Stewart2017, Wyatt+2018}. Initial conditions for these numerical simulation studies were created by first making a spherical planet, providing it with a small angular velocity, letting it relax in a smoothed particle hydrodynamical (SPH) simulation, then repeating this process until the desired angular velocity was reached \citep[supplementary materials]{Cuk+Stewart2012}. This method is slow and leads to pre-impact planets with structures that cannot be known until the end of this process.

The canonical impact model has been revived by the recent detection of oxygen isotope heterogeneity in returned lunar samples, where the signature of Theia becomes increasingly apparent in samples derived from deeper in the lunar mantle \citep{Cano+2020}. Of particular relevance for our study are the frequently sighted clumps of SPH particles amongst the post-impact debris \citep[e.g.][]{Benz+1987}. These clumps form in the tidal arm of debris coming from the part of Theia that does not directly strike the proto-Earth \citep{Canup2004a}. However, concerns over artificial clumping of SPH particles in shear flows \citep{Imaeda+2002} and numerical convergence of results during the chaotic post-impact evolution \citep{Canup+2013,Asphaug2014} have left uncertain whether or not this clump could be the proto-Moon. Even when clumps were present in high resolution SPH simulations, they were most notable for their effect on the material in the smoother debris disc because their orbits led them to collide with the Earth in a matter of hours \citep{Hosono+2017}.


Investigating the effect of spin on planetary collisions requires many simulations in which both the initial spin and internal composition are reliably generated. The challenge of finding an elegant and efficient way to construct initial conditions has significant overlap with attempts to model the internal structures of gas giant planets using their measured gravitational moments. Much of this work has made use of the concentric Maclaurin spheroid (CMS) method introduced by \citet{Hubbard2013}, where a uniformly spinning planet is described as a superposition of constant-density spheroids -- for which closed analytical solutions exist for the moments of the gravitational potential \citep{Hubbard2012}. This method has been developed to improve precision at high spin rates \citep{Kong+2013,Lock+Stewart2017}, include differential rotation on cylinders \citep{WisdomHubbard2016} and increase the number of spheroids that can be included for a given computing time \citep{Militzer+2019}. For example, measurements of the gravitational moments from the Juno mission \citep{Iess+2018} have been used in conjunction with CMS models to infer the extent of the differential rotation of Jupiter's atmosphere \citep{Guillot+2018}. The situation for Saturn is complicated by the presence of extensive rings, but the CMS method has also been used to analyse recent Cassini measurements and study the planet's more extensive differential rotation \citep{Iess+2019}.

In this paper, we present a fast algorithm that calculates the internal density profile of a rotating object composed of any prescribed materials in hydrostatic equilibrium, and places particles into the body such that very little, if any, relaxation is required for numerical simulations. The method is based on the CMS technique without differential rotation, but it allows arbitrary equations of state to be used for multiple material layers and exploits an analytical expression for the gravitational potential rather than using a slower and less accurate truncated expansion of Legendre polynomials. Our open-source code is a flexible tool that has been written in python under the project name WoMa (World Maker). It is described in~\cref{methods}, tested in~\cref{tests} and publicly available at \url{https://github.com/srbonilla/WoMa}. In~\cref{theia} we use WoMa to construct initial conditions for a set of giant impacts between the proto-Earth and Mars-sized impactors with a variety of rotation rates. Conclusions are presented in~\cref{conclusions}.

\section{Initial conditions generation}
\label{methods}

In this section, we describe our method for creating particulate realisations of uniformly spinning spheroids. It entails: (1) iteratively solving the equation of hydrostatic equilibrium to create an interior model of the spinning object, and (2) sampling the three-dimensional solution with particles, arranged such that their SPH densities match the desired values. 

\subsection{Interior model} 
\label{sec:hydro} 

Within the reference frame of a body spinning about its $z$-axis with constant angular velocity $\Omega$, the equation of hydrostatic equilibrium can be written using cylindrical coordinates, $\vec{r} = (r_{xy},\alpha,z)$, as
\begin{equation}
    \frac{1}{\rho}\nabla P=-\nabla \phi - \Omega^2 r_{xy}\,\widehat{r}_{xy} \; ,
	\label{eq:hydrorot}
\end{equation}
where $P(r_{xy},z)$, $\rho(r_{xy},z)$, and $\phi(r_{xy},z)$ represent the azimuthally symmetric pressure, density, and gravitational potential respectively. The third term represents the centrifugal force and is directed away from the rotation axis. The right hand side of equation~\autoref{eq:hydrorot} can be viewed as the negative gradient of an effective potential, $\Psi$, that includes the gravity and angular momentum barrier terms:
\begin{equation}
    \Psi = \phi + \frac{1}{2}\Omega^2 r_{xy}^2 \; .
	\label{eq:effpot}
\end{equation}

To solve equation~\autoref{eq:hydrorot} we also need a sufficient selection of the following quantities to make the problem well-defined: an equation of state (EoS, $P(\rho, T)$), and a temperature--density relation, $T(\rho)$, for each material; the pressure, $P_{\rm s}$, density, $\rho_{\rm s}$, and temperature, $T_{\rm s}$, at the surface of the object; the total mass, $M$, radius of the non-rotating solution, $R_{\Omega=0}$, and locations of any boundaries between distinct material layers in the non-rotating body, $R_{B,\Omega=0}$. Note that not all of these variables need to be specified as inputs for WoMa. The EoS, the temperature-density relation, and two of the three boundary conditions ($P_{\rm s}$, $\rho_{\rm s}$, and $T_{\rm s}$) must always be specified. However, various combinations of the other quantities can be used as inputs; for example WoMa can determine the total mass given the total radius for a one-layer object, or the boundary between materials given the total mass and radius for a body containing two distinct material layers. Many other combinations are available, particularly for three-layer planets.

The solution to equation~\autoref{eq:hydrorot} for a constant density object is the Maclaurin (oblate) spheroid \citep{Tassoul1978}, and more general solutions can be described as systems of overlapping concentric Maclaurin spheroids \citep{Hubbard2013}. Then, the density at any point inside the planet is the sum of the densities of all of the spheroids containing that point. As the isodensity surfaces are all spheroids, we can describe the full three-dimensional solution to equation~\autoref{eq:hydrorot} using just the equatorial and polar density profiles.

Our approach to solving equation~\autoref{eq:hydrorot} begins by finding the density profile for the spherically symmetric, non-rotating ($\Omega=0$) case. This solution, $\rho_{\Omega=0}(r)$, is evaluated in two  one-dimensional arrays, one in each of the equatorial and polar directions. These arrays both contain $N_g$ elements and span out to $r_{xy}=1.5R_{\Omega=0}$ and $z=1.2R_{\Omega=0}$. These maxima can be increased, if needed, for very rapidly rotating objects. 
$\rho_{i=0}(r_{xy},z)$ is used to compute the first value of the effective potential, $\Psi_1$, via 
\begin{equation}
    \Psi_i(r_{xy},z)=\iiint \frac{\rho_{i-1}(r_{xy},z)}{|\vec{r}-\vec{r}'|}d^3 r' + \frac{1}{2}\Omega^2 r_{xy}^2 \; .
	\label{eq:iterationV}
\end{equation}
The iteration loop is closed by updating the density, $\rho_i(r_{xy},z)$, as the solution to
\begin{equation}
    \frac{1}{\rho_i(r_{xy},z)}\nabla P_i(r_{xy},z) = -\nabla \Psi_i(r_{xy},z) \; ,
	\label{eq:iterationrho}
\end{equation}
where $P_i(r_{xy},z) = P(\rho_i, T_i)$ is determined by the EoS. These iterations do not conserve the total mass of the object. However, WoMa will loop over different $M_{\Omega=0}$ values, or whichever variable is relevant, until a solution is found with the desired mass.

The iterative process is continued until the mean fractional difference between the two last equatorial density profiles falls below a specified threshold. This corresponds to $c<10^{-3}$, where the convergence statistic is defined as
\begin{equation}
    c \equiv \frac{1}{N_g} \sum\limits_{j=1}^{N_g} \frac{|\rho_{i}(r_{xy, j}) - \rho_{i - 1}(r_{xy, j})|}{\rho_{i - 1}(r_{xy, j})},
    \label{eq:convergence}
\end{equation}
and the average is determined using only elements $j$ for which ${\rho_{i}(r_{xy, j})}$ and ${\rho_{i - 1}(r_{xy, j})}$ are both non-zero.

For uniform-density oblate spheroids, the three-dimensional integral to find the gravitational potential in equation~\autoref{eq:iterationV} can be recast as a one-dimensional integration. Defining the semi-major and semi-minor axes as $R$ and $Z$ respectively, the gravitational potential due to an oblate spheroid of constant density $\rho$, can be written as \citep{Kellogg1929}
\begin{equation}
    \phi(r_{xy},z)=-G\rho \pi R^2 Z \int_\lambda^\infty \left(1 - \frac{r_{xy}^2}{R^2+s} - \frac{z^2}{Z^2+s}\right)\frac{ds}{\sqrt{\varphi(s)}} \; ,
	\label{eq:kellogg}
\end{equation}
where $\varphi(s)\equiv(R^2+s)^2(Z^2+s)$, and $\lambda=0$ if $(r_{xy},z)$ lies within the spheroid, or the biggest root of the equation $f(s)=0$ otherwise, where

\begin{equation}
    f(s)\equiv\frac{r_{xy}^2}{R^2+s} + \frac{z^2}{Z^2+s} - 1 \; .
	\label{eq:fofs}
\end{equation}

Equation~\autoref{eq:kellogg} can be solved analytically along the axis $r_{xy}=0$ and in the plane $z=0$, both inside and outside the spheroid, to give
\begin{equation}
    \frac{\phi(z)}{A}=\left[\frac{2z^2}{\left(R^2 - Z^2\right)\sqrt{Z^2 + t}} + 2\left(\frac{R^2 + z^2 - Z^2}{\left(R^2 - Z^2\right)^{3/2}}\right)\tan^{-1}\gamma\right]_{t=\lambda}^{t\rightarrow\infty}
	\label{eq:kelloggintegral1}
\end{equation}
for $r_{xy}=0$, and 
\begin{equation}
\frac{\phi(r_{xy})}{A}=\left[\frac{-r_{xy}^2\sqrt{t + Z^2}}{\left(R^2 + t\right)\left(R^2 - Z^2\right)}+\frac{2R^2 + 2Z^2 - r_{xy}^2}{\left(R^2 - Z^2\right)^{3/2}}\tan^{-1}\gamma
    \right]_{t=\lambda}^{t\rightarrow\infty}
    \label{eq:kelloggintegral2}
\end{equation}
for $z=0$, where $A\equiv-G\rho\pi R^2Z$ and $\gamma^2\equiv(t + Z^2)(R^2 - Z^2)$. We use this analytical solution to calculate the gravitational potential rapidly at any point in space, rather than the usual truncated expansion in Legendre polynomials.

\subsection{Particle placement for SPH simulations}
\label{sec:picleplacement} 
In order to simulate a spinning body using a particle-based method like SPH, the solution for the density, $\rho(r_{xy},z)$, found in \cref{sec:hydro} must be converted into an appropriate set of particles. Desirable features of such a partitioning of the volume are that the particles should have very similar masses, no large-scale symmetries should be introduced that are not present in the body itself, and the particle distribution should be locally homogeneous to avoid introducing scatter in the densities of the particles. For spherically symmetric objects, these aims have led to approaches that place particles in nested spherical shells \citep{Saff+Kuijlaars1997, Raskin+Owen2016a, Reinhardt+Stadel2017, Kegerreis+2019}. 
In this subsection we build on the work of \citet{Kegerreis+2019}, generalising their stretched equal-area ({\sc SEA}) algorithm to the case where particles are placed into spheroidal isodensity shells.

\begin{figure}
\begin{tikzpicture}
\node[above right] (img) at (0,0) {\includegraphics[width=\columnwidth]{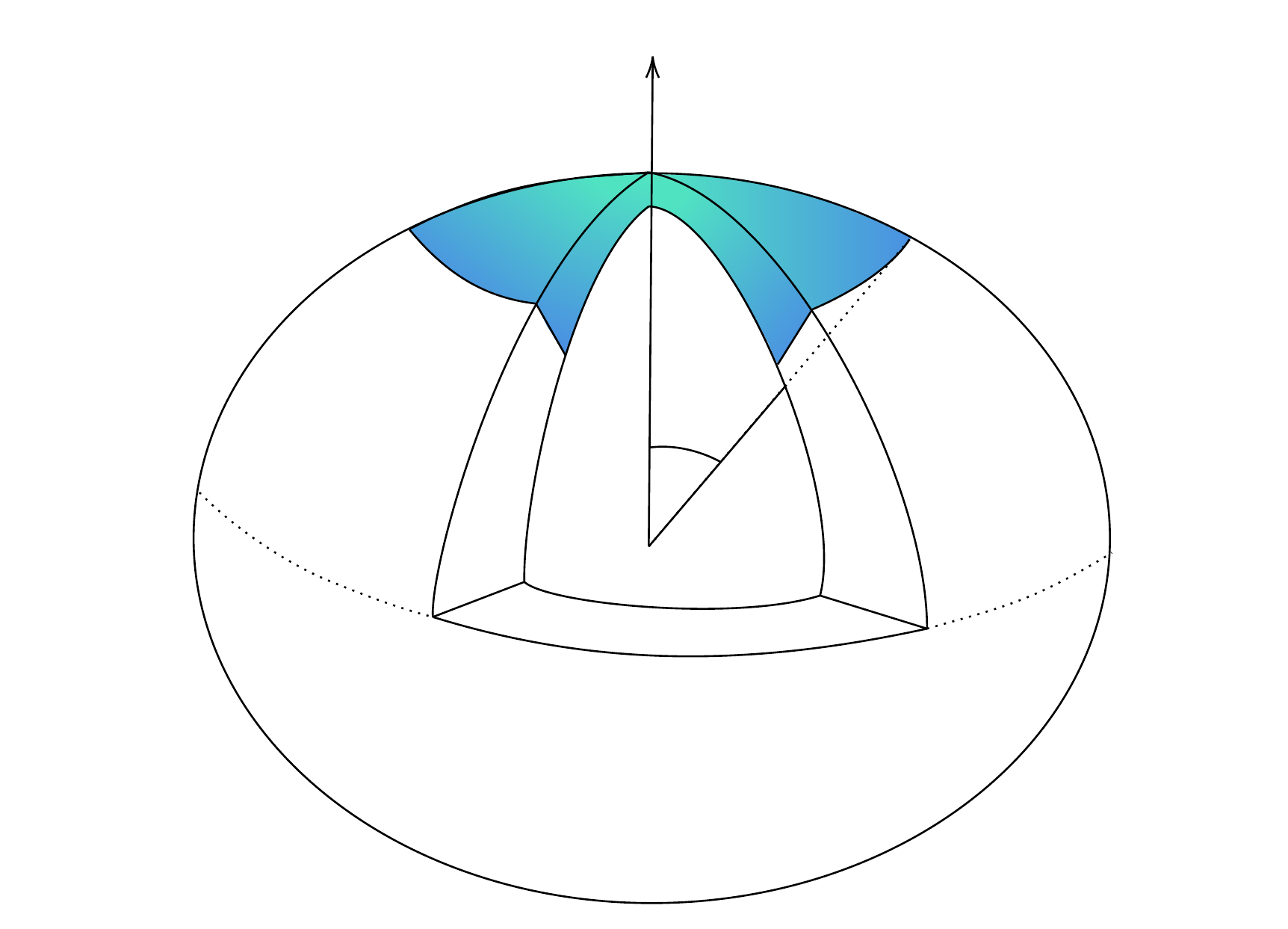}};
\node at (180pt,52pt) {$R_{\rm out}$};
\node at (150pt,70pt) {$R_{\rm in}$};
\node at (130pt,130pt) {$Z_{\rm in}$};
\node at (132pt,150pt) {$Z_{\rm out}$};
\node at (125pt,170pt) {$z$};
\node at (130pt,97pt) {$\theta$};
\end{tikzpicture}
	\vspace{-2em}
    \caption{Illustration of the fractional volume enclosed within a polar angle $\theta$, $V(<\theta)$, for a spheroidal shell, which dictates the latitudinal arrangement of particles required to represent a constant-density spheroidal shell.
    }
    \label{fig:spheroid}
    \vspace{-1em}
\end{figure}
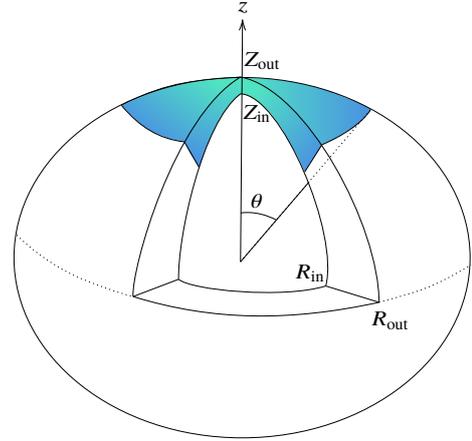

We start by using SEAGen \citep{Kegerreis+2019} to create a spherical object with a radial density profile matching the equatorial profile of our desired spheroid and containing the desired number of particles, $N$. SEAGen arranges $I$ spherical shells of particles such that the final one lines up with the edge of the body and any interior boundaries between different material layers are similarly accommodated. The midpoints of these shells have radii $R_i$ representing the semi-major axes of the shells of particles in our desired spheroidal object. The semi-major and semi-minor axis boundaries of the spheroidal shells are given by
\begin{equation}
\begin{split}
    R_{i,\rm out} &= \frac{R_i + R_{i+1}}{2}, \hspace{2mm}
    Z_{i,\rm out} = \frac{Z_i + Z_{i+1}}{2}, \\
    R_{i,\rm in} &= \frac{R_i + R_{i-1}}{2}, \hspace{2mm}
    Z_{i,\rm in} = \frac{Z_i + Z_{i-1}}{2}, \\
\end{split}
\end{equation}
where, by definition, $R_{i,\rm out} = R_{i+1,\rm in}$ and $Z_{i,\rm out} = Z_{i+1,\rm in}$. Using the solution to equation~\autoref{eq:hydrorot} calculated in \cref{sec:hydro}, the total mass in each spheroidal shell, $M_i$ can be computed. The number of particles in each spheroidal shell, $N_i$, is then set as the nearest integer to $(M_i/M)N$, to ensure that the total number of particles in the spheroid is as close as possible to the desired value $N$.

The SEAGen algorithm is employed again to create $I$ spherical shells that are randomly rotated with respect to one another, placing $N_i$ particles with mass $M_i/N_i$ into the $i^{\rm th}$ shell. To transform from spherical to spheroidal shells, each particle is: (1) shifted in polar angle, $\theta$, to reproduce the cumulative mass (or equivalently, volume or particle number) fraction distribution of the spheroidal shell, $f_i(<\theta)$, then (2) mapped at fixed polar and azimuthal angle to place it onto the required spheroidal shell.

SEAGen provides us with spherical isodensity shells of particles, which have a cumulative fractional number that satisfies
\begin{equation}
    f_{\rm sphere}(<\theta)=\left(1 - \cos\theta\right)/2.
\end{equation}
The corresponding function for the $i^{\rm th}$ isodensity spheroidal shell, $f_i(< \theta)$, is more complicated because the shell has a $\theta$-dependent radius and thickness.
For the $i^{\rm th}$ spheroidal shell, bounded by the spheroids with semi-major and semi-minor axis pairs ($R_{i,{\rm in}}$, $Z_{i,{\rm in}}$) and ($R_{i,{\rm out}}$, $Z_{i, {\rm out}}$), the cumulative enclosed volume, as illustrated in~\autoref{fig:spheroid}, can be written as
\begin{equation}
\begin{split}
    V_i(<\theta) &= \int_{0}^{2\pi}d\phi' \int_{0}^{\theta}\sin\theta' \int_{r_{ i,{\rm in}}(\theta')}^{r_{i,{\rm out}}(\theta')}{r'}^2 dr' d\theta' \\
    &= \frac{2\pi}{3}\int_{0}^{\theta}\left(r_{i,{\rm out}}(\theta')^3 - r_{i, {\rm in}}(\theta')^3\right)\sin\theta' d\theta'\,,
	\label{eq:Vthetaintegral}
\end{split}
\end{equation}
where $r_{i,j}(\theta') = \left(\dfrac{\sin^2(\theta')}{R_{i,j}^2} + \dfrac{\cos^2(\theta')}{Z_{i,j}^2}\right)^{-1/2}$, with ${j=\{\rm in\,,\rm out\}}$, is the distance from the centre of the coordinate system to the inner or outer spheroid surface, at a given polar angle. 
The solution of~\autoref{eq:Vthetaintegral} is 
\begin{equation}
\begin{split}
    V_i(<\theta) = \dfrac{2\pi}{3}&\left[\sqrt{2}\left(-\dfrac{R_{i,{\rm out}}^2}{F_{i,{\rm out}}(\theta)} 
    + \dfrac{R_{i,{\rm in}}^2}{F_{i,{\rm in}}(\theta)}\right)\cos{\theta}\right. \\
    &+\left.R_{i,{\rm out}}^2 Z_{i,{\rm out}} - R^2_{i,{\rm in}} Z_{i,{\rm in}}\right] \;, 
\end{split}
\end{equation}
where
\begin{equation*}
    F_{i,j}(\theta) = \sqrt{R_{i,j}^{-2} + Z_{i,j}^{-2} + \left(-R_{i,j}^{-2} + Z_{i,j}^{-2}\right)\cos2\theta} \;. \\
\end{equation*}
As the shell is assumed to have a uniform density, we can infer that the cumulative fractional number of particles in the spheroidal shell should satisfy ${f_i(<\theta)=V_i(<\theta)/V_i(<\pi)}$. 

Having determined $f_i(<\theta)$ for each shell and $f_{\rm sphere}(<\theta)$, we can now define the polar angle mapping of the particles on a SEAGen-generated spherical shell to the corresponding spheroidal shell via
\begin{equation}
    \theta \rightarrow f_i^{-1} \left(f_{\rm sphere}(<\theta)\right)=f_i^{-1}\left(\frac{1 - \cos\theta}{2}\right).
\end{equation}
With the particles now distributed in a uniform and unbiased way with respect to the polar angle, the final step is to map their radial positions from the spherical shell that SEAGen placed them on to the desired spheroidal one, using
\begin{equation}
    r\rightarrow \left(\dfrac{\sin^2(\theta)}{R_i^2} + \dfrac{\cos^2(\theta)}{Z_i^2}\right)^{-1/2} r.
\end{equation}

\section{Tests of the initial conditions generation}
\label{tests}

In this section we test the WoMa algorithm described in~\cref{methods}, in particular the iterative method to solve the equation of hydrostatic equilibrium for a uniformly rotating spheroid, and the technique to distribute particles to produce a low-noise representation of the solution. To demonstrate the capabilities of WoMa, we construct one- and two-layer planets using from $10^5$ particles -- as are commonly used in SPH simulations of planetary giant impacts -- up to $10^9$ particles, an order of magnitude more than the highest numbers to date \citep{Hosono+2017,Kegerreis+2019,Kegerreis+2020}, and evolve them to check how relaxed these initial conditions actually are.

\subsection{Finding the dynamical equilibrium configuration} 
\label{sec:hydrotest} 

\begin{table}
    \begin{center}
    \caption{Properties of the planets used as input to WoMa's iterative solution of~\autoref{eq:hydrorot} for the one- and two-layer test cases.}
	\label{tab:sphericalcond}
    \begin{tabular}{ |l|l|c|c| } 
        \hline
        Property &&One-layer&Two-layer\\
        \hline
        Mass&$M$ [M$_\oplus$] & 0.640 & 1 \\ 
        Radius&$R_{\Omega=0}$ [R$_\oplus$] & 1 & 1 \\
        Boundary&$R_{B,\Omega=0}$ [R$_\oplus$]& - & 0.481 \\ 
        Period&$T_{\Omega}$ [hours]& 3.25 & 2.60 \\ 
        Surface density&$\rho_{\rm s}$ [kg~m$^{-3}$]& 2450.1 & 2511.8 \\ 
        Surface pressure&$P_{\rm s}$ [Pa] & $10^5$ & $10^5$ \\ 
        Surface temperature&$T_{\rm s}$ [K] & 3000 & 300 \\
        \hline
    \end{tabular}
    \end{center}
\end{table}

\begin{figure}
	\includegraphics[width=\columnwidth]{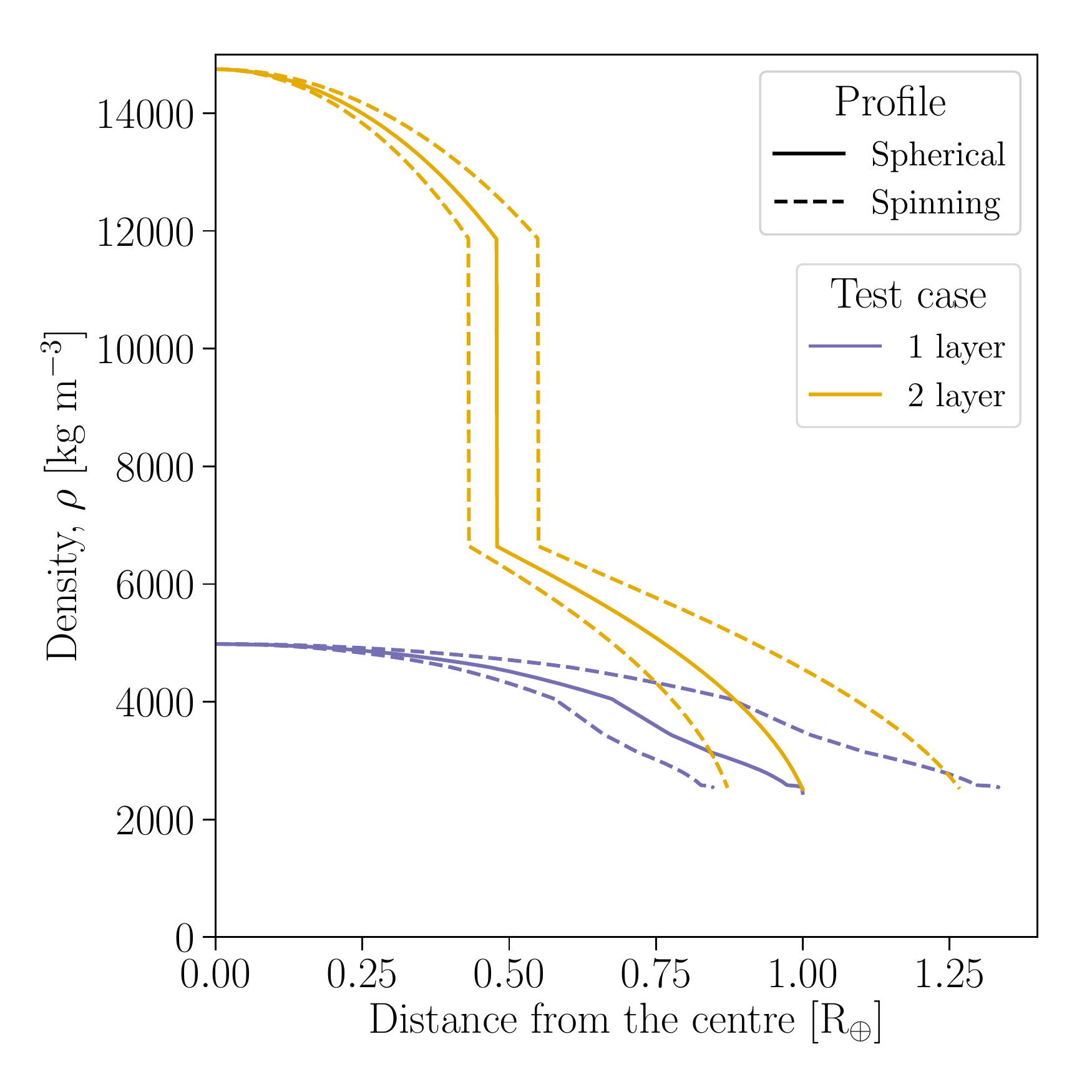}
	\vspace{-2em}
    \caption{Analytically--solved density profiles for the spherical and spinning planets. Solid lines represent the initial spherical models, and dashed lines represent the equatorial and polar density profiles of the corresponding uniformly rotating fluid planets that solve equation~\autoref{eq:hydrorot}. Note that the kinks in the profile of the one-layer planet reflect features in the SESAME basalt EoS.}
    \label{fig:spinning_profiles}
    \vspace{-1em}
\end{figure}

\begin{figure}
	\includegraphics[width=\columnwidth]{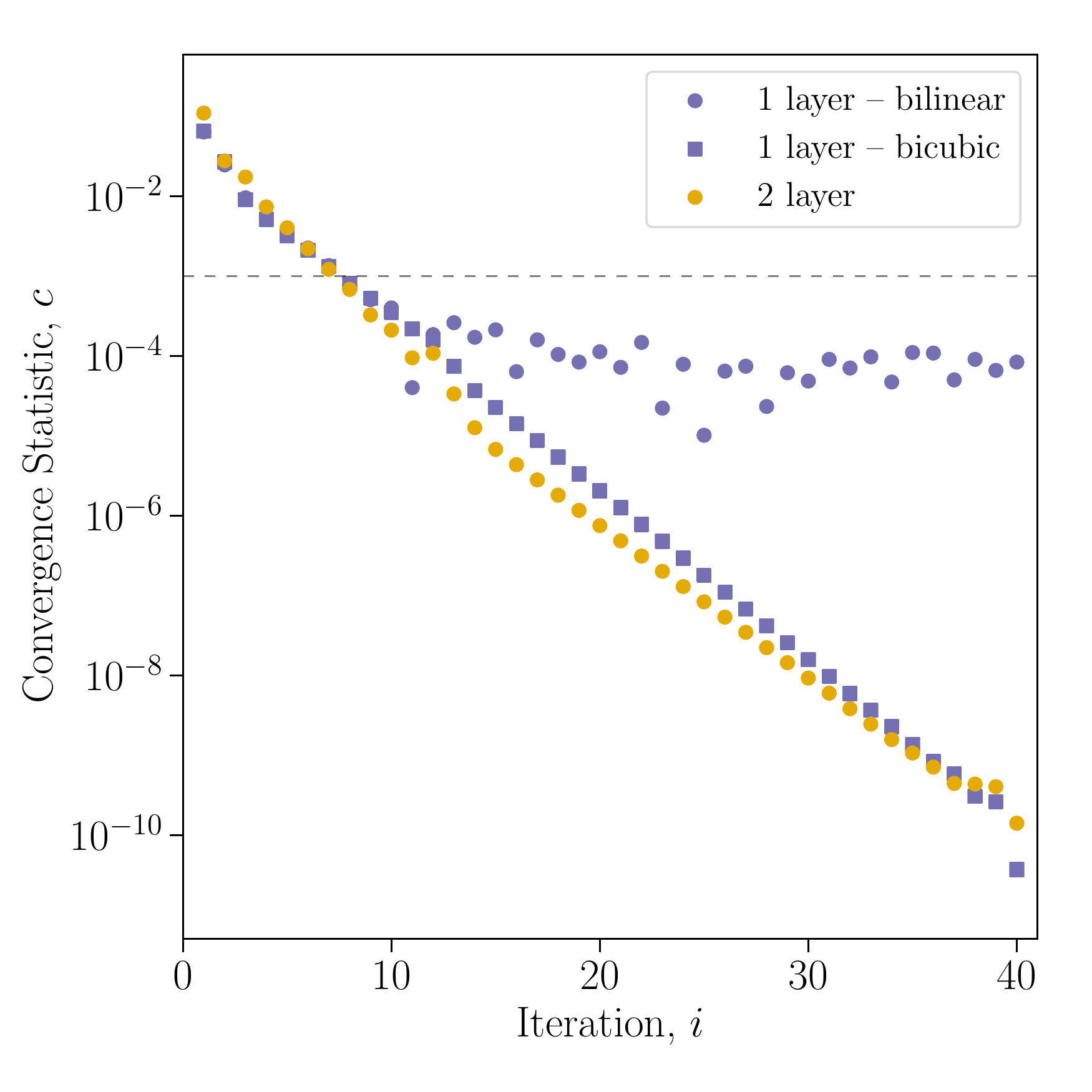}
	\vspace{-2em}
    \caption{Convergence of the equatorial density profile, measured by the statistic $c$ defined in equation~\autoref{eq:convergence}, as a function of iteration number.
    For the one-layer test we use two versions of the basalt EoS with either bilinear or bicubic interpolation of the SESAME tables. Bicubic interpolation ensures that the derivatives of $P(\rho,T)$ are continuous at every given $\rho$ and $T$ and produces better convergence. $c < 10^{-3}$, beneath the dotted line, is more than sufficient for making SPH initial conditions such as those we use here, and is rapidly achieved by WoMa.}
    \label{fig:iteration_e}
    \vspace{-1em}
\end{figure}

WoMa has been written in a modular way such that different EoS and temperature-density relations can be readily included. For our test planets we use publicly available EoS:
SESAME basalt \citep{SESAME} to describe the material comprising the one-layer planet, and Tillotson iron and granite \citep{Tillotson1962, Melosh1989} for the two-layer planet core and mantle respectively. The one-layer planet is assumed to be isothermal, whereas  $T=k\rho^{2.5}$ is chosen for illustrative purposes for both materials of the two-layer test case. The other parameters and boundary conditions describing the non-spinning objects, from which WoMa iterates to find the rotating bodies, are given in~\autoref{tab:sphericalcond}. ~\autoref{fig:spinning_profiles} shows, with solid lines, the radial density profiles for the resulting spherically symmetric solutions.

We define the maximally-spinning body to be the most rapidly rotating one for which the centrifugal force does not overcome gravity at any point within the object. If this requirement is violated within our uniformly rotating body, then the force resulting from the pressure gradient would need to act inwards, leading to an unphysical situation. For our solid body rotating one- and two-layer planets, these maximum spins correspond to periods of $T_{\Omega,{\rm min}}=3.03$ and $2.27$ hours respectively. We choose very short periods of $T_\Omega=3.25$ hours for the one-layer test, and $T_\Omega=2.60$ hours for the two-layer case, in order to yield significantly flattened objects. Using grids with $N_g=10^5$ elements, the iterative procedure within WoMa finds the equilibrium configurations for our test planets with their respective rotation periods. The polar and equatorial density profiles are shown, for both cases, using dashed lines in~\autoref{fig:spinning_profiles}.

It is important to demonstrate the convergence of this iterative scheme. To this end, we use the convergence statistic defined in equation~\autoref{eq:convergence}, applied to
the $N_g$-sized equatorial density array. This is simply computing the mean fractional change in the equatorial density profile over the previous iteration. The evolution with iteration number of the convergence statistic is shown in~\autoref{fig:iteration_e}. For the purposes of typical numerical simulations, the precision necessary is reached in only a few iterations, with both the one-and two-layer cases able to converge to much higher precision provided that a differentiable EoS is employed. For a bilinear interpolation of the SESAME basalt EoS table,~\autoref{fig:iteration_e} shows that the convergence reaches a floor, albeit one in this instance that is still sufficiently low for our purposes. The height of this floor depends somewhat on the coarseness of the computational grid. For the results shown here, $N_g=10^5$, and 15 iterations take under 5~minutes to compute on a common desktop computer. This resolution will suffice for the mass of the innermost spheroid to be smaller than the mass of one particle when placing up to $\sim 10^{11}$ particles.

\subsection{Particle placement for SPH simulations}
\label{sec:picleplacementtest} 

\begin{figure*}
	\hspace*{-1cm} 
	\includegraphics[width=5.5in]{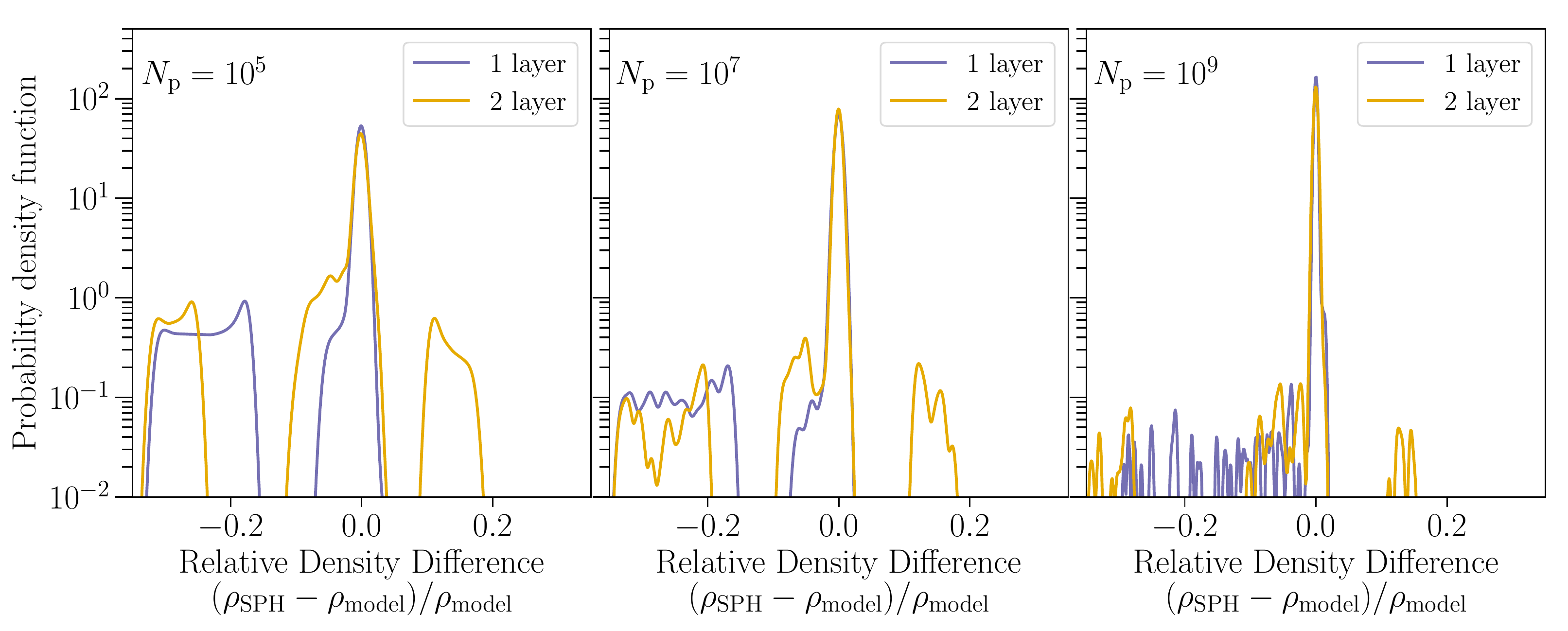}
	\vspace{-0.5em}
    \caption{Comparison of our model solution for hydrostatic equilibrium with the SPH density computed at each particle's location in the initial conditions. The first, second, and third columns contain the initial conditions with $10^5$, $10^7$, and $10^9$ particles respectively, and the different colours reflect the results for the one- and two-layer planets.}
    \label{fig:simulation_t0}
    \vspace{-1em}
\end{figure*}

To test that the particle placement algorithm leads to low-noise representations of the test planets we use WoMa to create particle representations of both cases using $10^5$, $10^7$, and $10^9$ particles. We test the accuracy of these representations by computing the particles' smoothed densities using the SPH code SWIFT \citep[\href{http://www.swiftsim.com}{www.swiftsim.com},][]{Schaller+2016} and 45 nearest neighbours. The distributions of fractional density errors are shown in~\autoref{fig:simulation_t0}. These distributions are all sharply peaked around zero, with full-widths at half-maximum of the peaks of $0.015$, $0.012$, and $0.005$ for the $10^5$, $10^7$, and $10^9$ particle realisations respectively. Better numerical resolution means decreased SPH smoothing lengths that sample a smaller range of densities, and lower stochastic errors in the sampled densities.

In addition to the bulk of the particles that lie in the central peak of the density error distribution,~\autoref{fig:simulation_t0} shows some particles whose densities differ by up to 40\%. These outliers arise at the density discontinuities of the outer surface and inner boundary between materials, and are an unavoidable consequence of how the standard SPH formulation computes densities by averaging over a number of nearby neighbours -- a well-known issue when performing SPH computations of a density profile with discontinuities \citep{Woolfson2007, Reinhardt+Stadel2017}. Our choice of SPH formulation with smoothed densities is entirely responsible for this part of the error distribution, not the particle placement being performed by WoMa. ~\autoref{fig:simulation_t0} shows that using more particles decreases the fraction of density outliers, because in these cases a smaller fraction of the particles lie near to boundaries.

\begin{figure*}
	\hspace*{-1cm} 
	\includegraphics[width=7.5in]{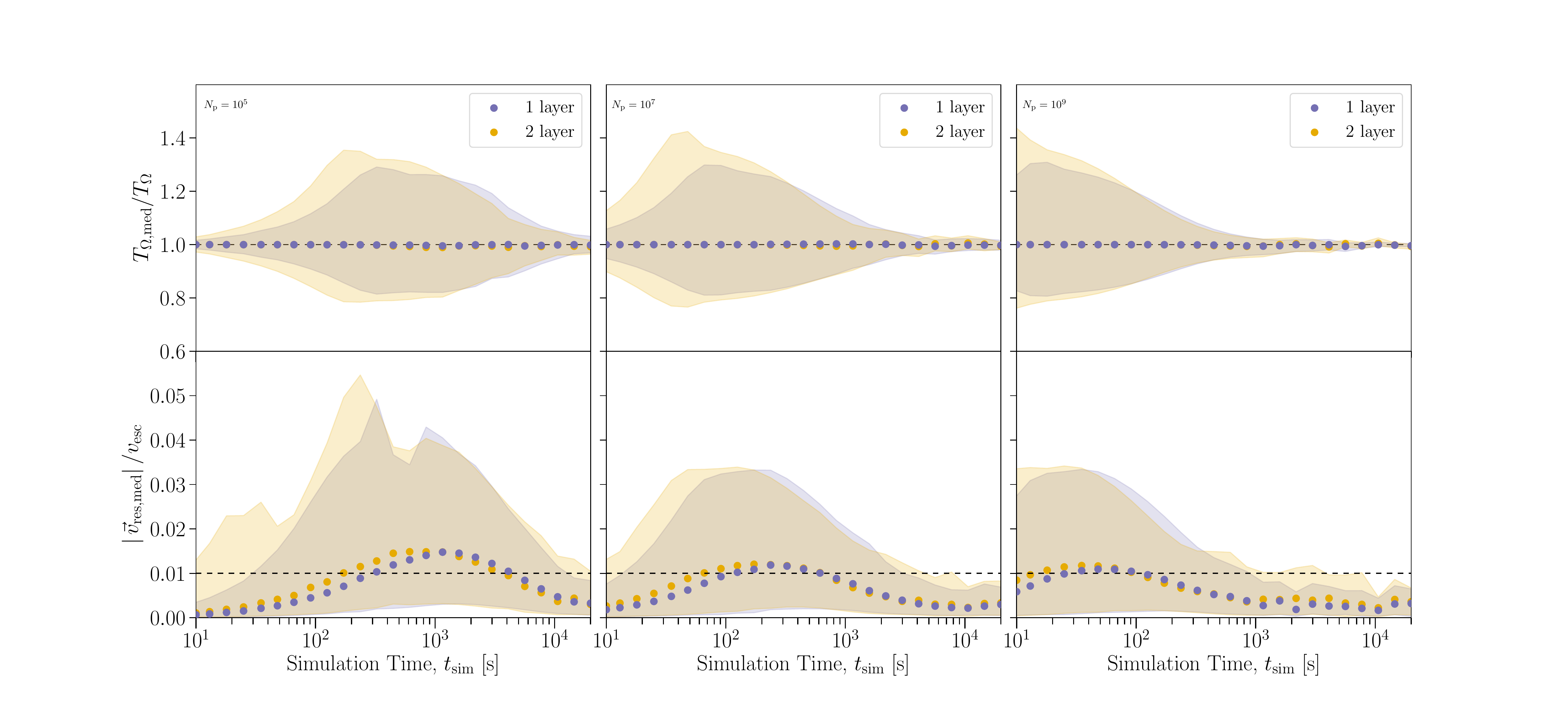}
	\vspace{-1.9em}
    \caption{Evolution of the median normalised period $T_{\Omega,{\rm med}}/T_{\Omega}$ (upper row), and median normalised residual speed $\left|\,\vec{v}_{\rm res, med}\right|/v_{\rm esc}$ (lower row). Shaded regions represent the 1st and 99th percentiles of the distributions. The first, second, and third columns contain the simulations with $10^5$, $10^7$, and $10^9$ particles respectively. A horizontal line shows where the residual velocity has a magnitude that is 1\% of the escape speed, a criterion sometimes used to define when initial conditions are relaxed.}
    \label{fig:simulation_evolution}
    \vspace{-1em}
\end{figure*}

We now test how close to equilibrium our test planets are by using the SWIFT code to evolve them both, for the three resolution levels, in a non-rotating reference frame. Each isolated rotating body is evolved to a simulation time of 20,000 s, i.e. just over 5.5 hours. This is close to 2 full rotations for both planets, and is several times the time taken for a sound wave to traverse the planet, so will be long enough to detect signs of disequilibrium.
We measure a rotation period and a residual velocity for each particle. These are calculated as
\begin{equation}
\begin{split}
    T_{\Omega,i}&=\frac{2\pi}{\Omega_i}\;, \quad \text{where }\quad \Omega_i=\frac{v_{\alpha,i}}{\sqrt{x_i^2 + y_i^2}} \;, \quad \text{and}\\
    \vec{v}_{\mathrm{res}, i}&=\vec{v}_{i} - \vec{\Omega} \times \vec{r}_i\; ,
\end{split}
\end{equation}
i.e. the SPH velocity minus the velocity each particle should have according to its position and the chosen angular velocity. The evolution of the median of the particle rotation periods, normalised by the desired period, and the median residual speed relative to the escape speed are shown in~\autoref{fig:simulation_evolution}, along with the 1$^{\rm st}$ and 99$^{\rm th}$ percentiles. The distribution of particle periods has a median that matches the desired value to within 1\% at all times, and very low scatter by the end of the simulations in all cases. Median residual speeds never reach 2\% of the escape speed, and barely exceed the 1\% level for the two higher resolutions. We define a set of initial conditions as ``relaxed" when the median particle speed is below 1\% of the escape speed. In both the one- and two-layer cases, higher resolutions lead to shorter relaxation times. 

We also compute the fractional density error distribution at the end of the simulation. Full-widths at half-maximum of the peaks are at $0.007$, $0.004$, and $0.006$ for the one-layer test case, and $0.016$, $0.008$, and $0.019$ for the two-layer test case, for the $10^5$, $10^7$, and $10^9$ particle realisations respectively. The final density profiles, excluding the boundaries, are within 2\% of the desired, analytically--computed density for the one- and two-layer tests with $10^5$ particles, and within 1\% for the two higher resolutions.

\section{The effects of a spinning Theia}
\label{theia}

\begin{figure*}
	\includegraphics[trim=150 40 150 50, clip,width=\textwidth]{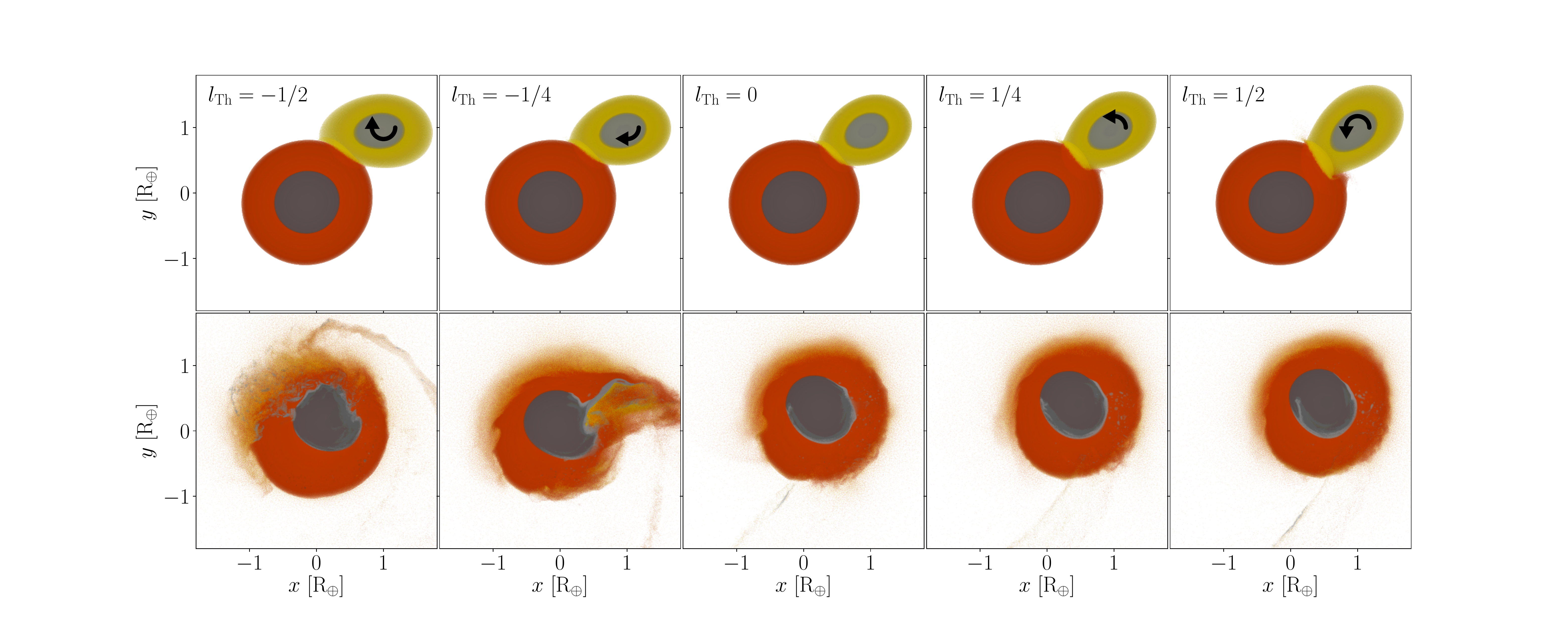}
    \caption{Snapshots in the early stages of the five giant impacts. Rows represent simulation times of 1 hour (top) and 5 hours (bottom). Columns represent different simulations with $l_{\rm Th}=-\tfrac{1}{2}$, $-\tfrac{1}{4}$, $0$, $\tfrac{1}{4}$, and $\tfrac{1}{2}$. 50 slices of thickness $\Delta z=0.12~R_\oplus$ are plotted in order of increasing $z$ from $z=-6~R_\oplus$ to $z=0$, which lies in the plane containing the centre of mass. The particle colours represent different materials: dark and light grey for Tillotson iron, and red and yellow are Tillotson granite, in the proto-Earth and Theia respectively. The origin of the coordinate system is taken to be the centre of mass of all the material in the simulation.}
    \label{fig:snapshots}
    \vspace{-1em}
\end{figure*}

\begin{figure*}
	\includegraphics[trim=50 40 30 40,clip, width=\textwidth]{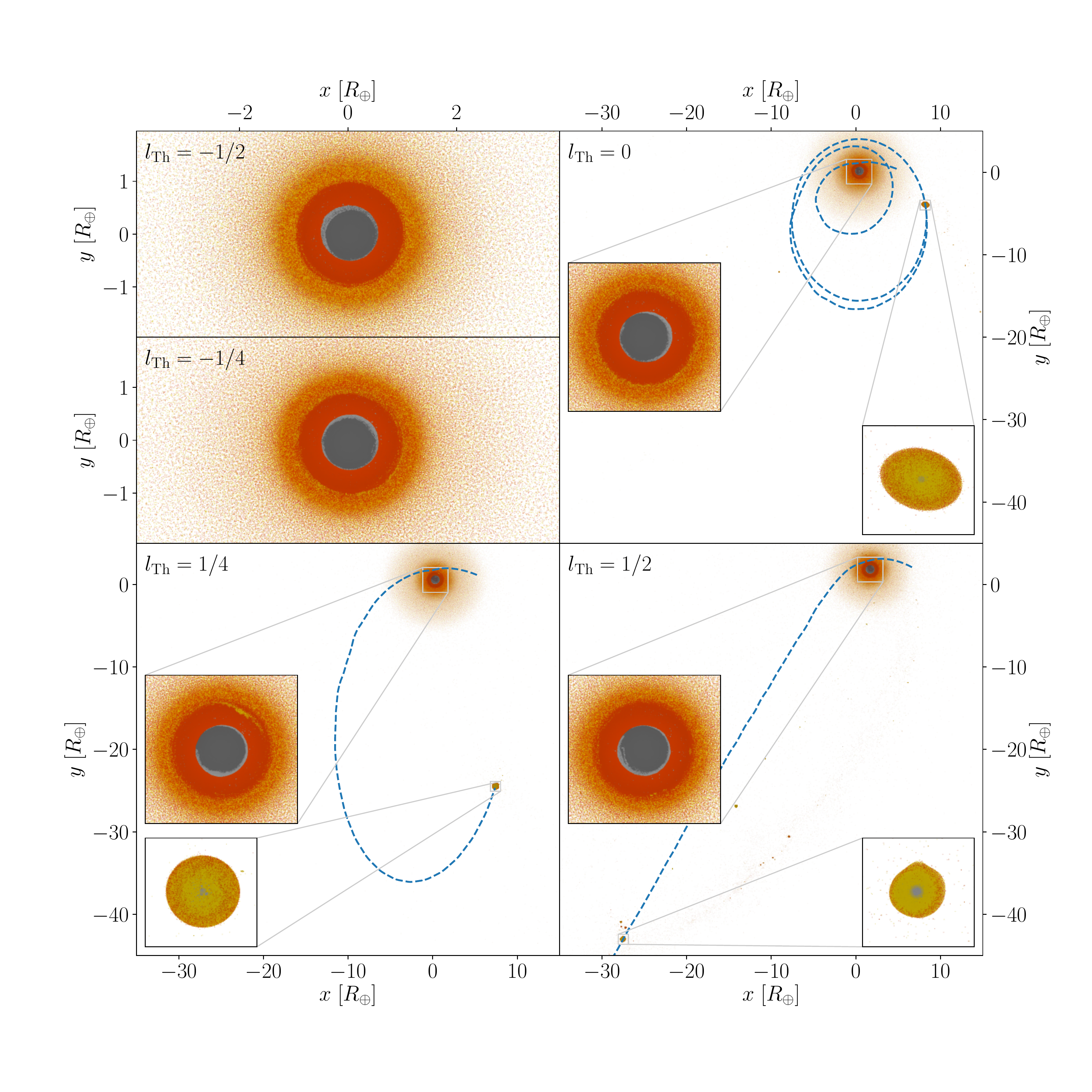}
	\vspace{-1.9em}
    \caption{Snapshots at 100 hours of simulation time for simulations with $l_{\rm Th}=-\tfrac{1}{2}$, $-\tfrac{1}{4}$, $0$, $\tfrac{1}{4}$, and 40 hours for the $l_{\rm Th}=\tfrac{1}{2}$ simulation. Blue lines represent the trajectories of the resulting clumps, and the particle colours are the same as in~\autoref{fig:simulation_t0}. The inset boxes show regions of side length $3~R_\oplus$ centred on the target and $1.2~R_\oplus$ for the clump in each panel. An animation of the early evolution of these impacts is available at \href{http://icc.dur.ac.uk/giant_impacts/woma_impacts_anim.mp4}{icc.dur.ac.uk/giant\_impacts}.}
    \label{fig:snapshots2}
    \vspace{-1em}
\end{figure*}

In this section, we present a set of five canonical Moon-forming giant impacts where the impactor Theia is given a different spin in each simulation. The full parameter space for such a study includes the spin angular velocity vectors of both proto-Earth and Theia, but in this initial study we restrict ourselves to situations with the spin and orbital angular momenta either parallel or anti-parallel and the target not initially rotating. Collisions between rotating protoplanets have been considered previously, both for the Moon-forming collision \citep{Canup2008, Cuk+Stewart2012, Nakajima+2015, Wyatt+2018} and terrestrial planets more generally \citep{Timpe+2020}, but not at particularly high numerical resolution. Recent studies \citep{Hosono+2017, Kegerreis+2019} have shown that at least $10^7$ particles can be required to converge on even large-scale results. The combination of WoMa and SWIFT enable us to produce better resolved simulations to investigate how Theia's spin can alter the outcome of the canonical Moon-forming collision.

We consider an impact between a target proto-Earth of mass 0.887~$M_\oplus$ and an impactor, Theia, of mass 0.133~$M_\oplus$. Both are differentiated into an iron core and rocky mantle, constituting 30\% and 70\% of the total mass respectively modelled using the \citet{Tillotson1962} iron and granite equations of state.
The Tillotson EoS is widely used for SPH impact simulations due to its computationally convenient analytical form \citep{Stewart2019}. However it does not treat phase boundaries or mixed phases correctly. Since the focus of this paper is the overall range of outcomes due to the spin of an impactor, the details of the EoS are not expected to have a significant affect on the main results.
The velocity at impact is chosen to be the mutual escape speed, the angle of impact is set as 45$^{\circ}$, and the simulation begins 1 hour prior to the time of contact between the two bodies in order to model the tidal distortion of the bodies just before impact. 
We give the iron and granite layers a temperature--density relation of $T\propto\rho^2$. With a 500~K surface temperature on both bodies, this yields a core temperature for the proto-Earth of $\sim 5000$~K, similar to the Earth today. All 5 simulations are evolved to 100 hours, and have a mass resolution of $10^7$ particles per Earth mass.

The only difference between our simulations is the rotation rate of Theia. The minimum period available is 2.6 hours, which translates to a maximum spin angular momentum of $L_{\rm Th,\rm max}=0.15$ $L_{\rm EM}$, where $L_{\rm EM}=3.5\times 10^{34}$~kg~m$^2$~s$^{-1}$ is the current angular momentum of the Earth--Moon system. We set the spin angular momentum of Theia, $L_{\rm Th}$, to be $l_{\rm Th} \equiv L_{\rm Th}/L_{\rm Th,max}=-\tfrac{1}{2}$, $-\tfrac{1}{4}$, $0$, $\tfrac{1}{4}$, and $\tfrac{1}{2}$ for our five simulations. These correspond to rotation periods for the more and less rapidly spinning Theias of 3.2 and 5.1~hours. The orbital angular momentum of the colliding systems is $1.25\,L_{\rm EM}$, which is only $\sim 0.05\,L_{\rm EM}$ larger than the values of the successful canonical impacts found by \citep{Canup+2001}.

~\autoref{fig:snapshots} shows cross sections of the moment of contact and a snapshot 4 hours later for each of the simulations. The most striking feature at the moment of impact is the difference in Theia's tidal distortion, which would have been absent had we started the simulation at the point of contact, as is often done in planetary giant impact studies. The tidal bulge of the more rapidly counter-rotating impactor ($l_{\rm Th}=-\tfrac{1}{2}$) has spun significantly ahead of the line joining the centres of the proto-Earth and Theia, stretching Theia along its direction of motion. For the rapidly co-rotating case ($l_{\rm Th}=\tfrac{1}{2}$), Theia's spin shifts the tidal distortion to point along the line connecting the centres of the bodies. Four hours after first contact, the counter-rotating impacts are near to completing their mergers, whereas the non-spinning and co-rotating largest remaining objects have drifted away from the origin of the coordinate system, chosen as the centre of mass of the simulation, reflecting the presence of significant unaccreted mass. The impactor core and mantle material that has already been deposited into the target is found towards the edges of the corresponding layers of the target, with more mixing between the two mantles. 

\begin{figure}
	\includegraphics[width=\columnwidth]{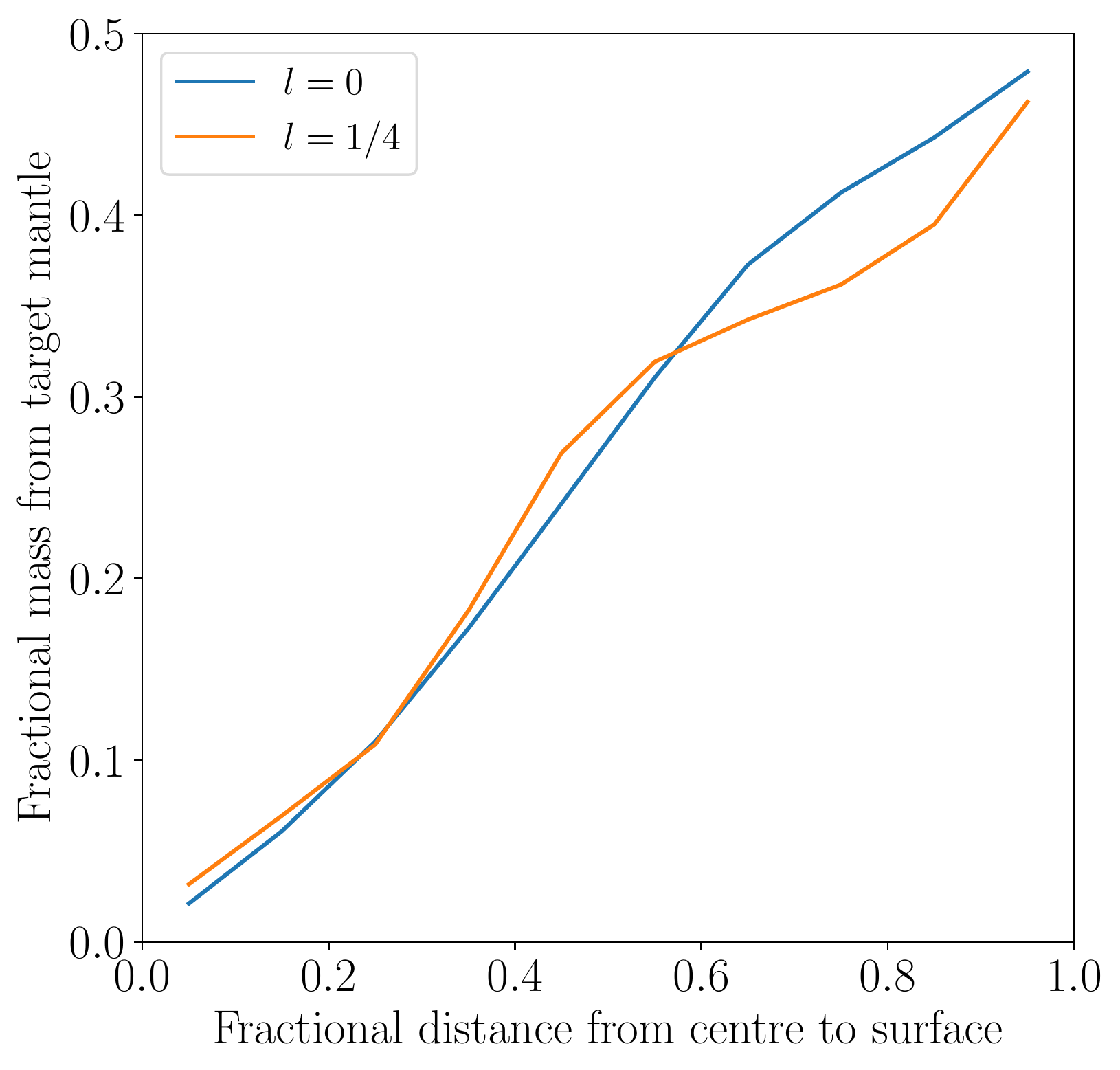}
	\vspace{-2em}
    \caption{Radial variation of the mass fraction of target mantle present in the orbiting clumps after 100 hours, for the $l=0$ (blue) and $l=\frac{1}{4}$ (orange) simulations.}
    \label{fig:composition_clumps}
    \vspace{-1em}
\end{figure}

\begin{figure}
	\includegraphics[width=\columnwidth]{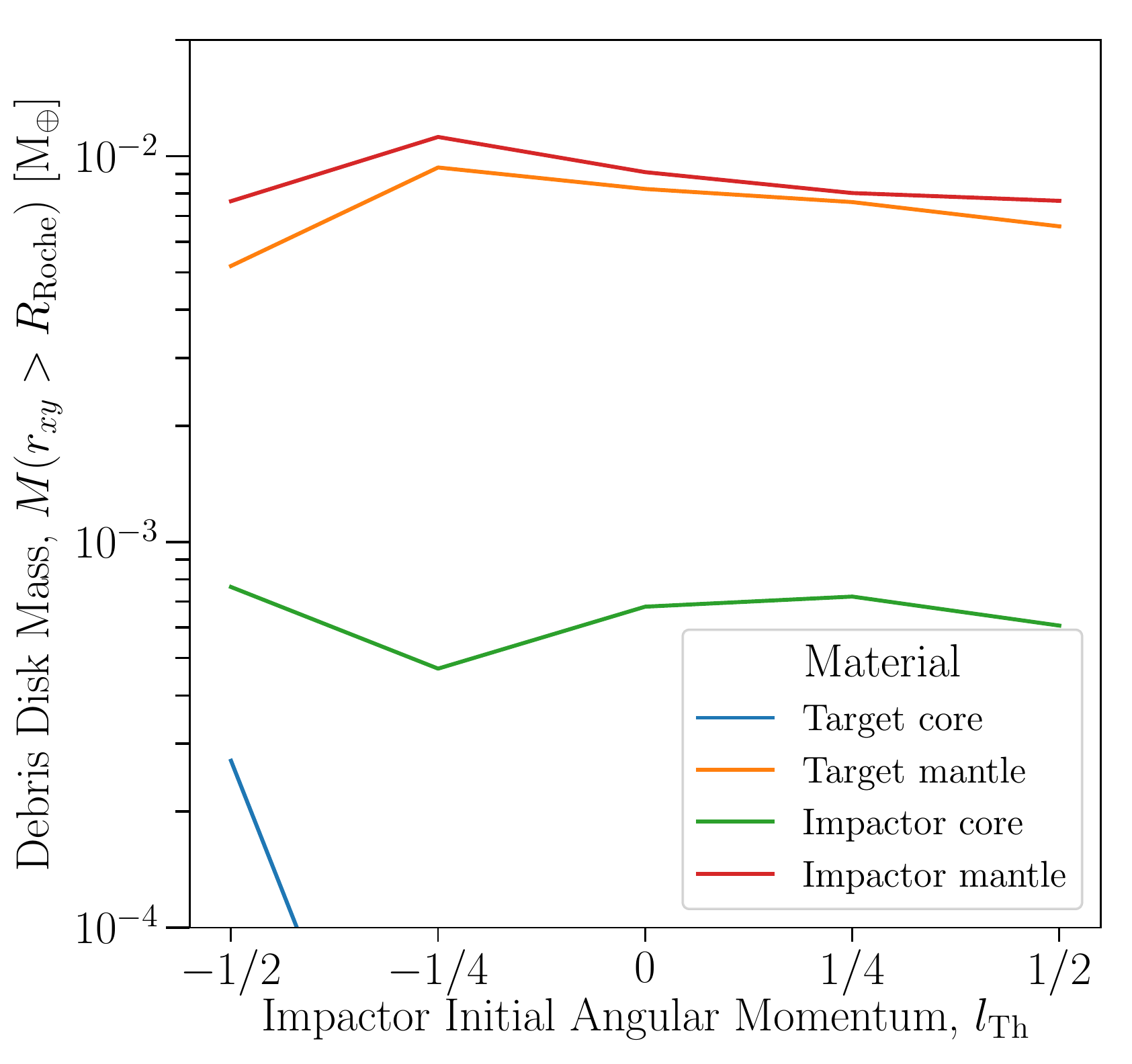}
	\vspace{-2em}
    \caption{Total bound mass, excluding clumps, outside 
    a cylinder with the Roche radius centred on the post-impact Earth, as a function of Theia's initial spin angular momentum. The four different colours represent the different material components as described in the legend. All results are at a simulation time of 100 hours.
    }
    \label{fig:composition}
    \vspace{-1em}
\end{figure}

Later snapshots of the five simulations are shown in~\autoref{fig:snapshots2}. Snapshots are shown at 100 hours of simulation time except for the $l_{\rm Th}=\tfrac{1}{2}$ case, which is at 40 hours to capture the highlighted clump before it flies out of frame. All $l_{\rm Th}<\tfrac{1}{2}$ simulations place the majority of the mass of Theia either into the Earth or within the Roche radius ($\sim 3~R_\oplus$), with much of Theia's core blanketing that of the proto-Earth. However, the $l_{\rm Th}=0$ and $\tfrac{1}{4}$ impacts lead to a large, self-gravitating clump within the debris disc that extends beyond the Roche radius. The formation of this clump is sensitive to the initial spin of Theia, because both counter-rotating impactor scenarios end in mergers and the rapidly corotating Theia produces a hit-and-run collision. In order of increasing $l_{\rm Th}$, the final gravitationally bound mass is $0.988$, $0.987$, $0.987$, $0.985$, and $0.955$ $M_\oplus$ out of the total of 1.020~$M_\oplus$. The corresponding total angular momenta, in units of $L_{\rm EM}$, evolve as follows: $1.17\rightarrow1.17, 1.21\rightarrow1.19, 1.25\rightarrow1.21, 1.28\rightarrow1.23$ and $1.32\rightarrow0.78$. In the $l_{\rm Th}=\tfrac{1}{2}$ case, the largest of the many escaping clumps has a mass of $0.0076~M_\oplus$, is taking away $0.047~L_{\rm EM}$ of the angular momentum and is $53.5~R_\oplus$ away from the Earth at 40 hours, beyond the edge of the region shown in the lower right-hand panel of~\autoref{fig:snapshots2}.

The clumps left orbiting the Earth after impact in the simulations with $l_{\rm Th}=0$ and $\tfrac{1}{4}$ have masses of 0.010 and 0.020~$M_\oplus$ respectively -- $0.813$ and $1.626$ times the mass of the present-day Moon. Their orbits have periods of 40 and 120 hours, eccentricities of 0.6 and 0.8, periapses of $\sim 4.6~R_\oplus$ and $2.8~R_\oplus$, and spin periods of 10 and 40 hours respectively. The periapse of the $l_{\rm Th}=0$ clump is well outside the Roche radius, so this proto-Moon, while enduring tidal distortion as shown in~\autoref{fig:snapshots2}, should survive. For the $l_{\rm Th}=\tfrac{1}{4}$ clump, the periapse lies just within the Roche radius of $3~R_\oplus$, but if the orbit circularises in the same way as can be seen for the $l_{\rm Th}=0$ proto-Moon, then this may enable a portion of it also to survive as a coherent proto-Moon.

The orbiting clump is resolved with over $10^5$ particles in the $l_{\rm Th}=0$ and $\tfrac{1}{4}$ simulations, allowing us to study in detail its composition. Both clumps have $\sim29\%$ of their mass coming from the proto-Earth's mantle, $\sim1\%$ from Theia's iron core, and the remaining $\sim70\%$ from Theia's mantle. No iron was present in the long-lived clumps found in the study of~\citet{Canup2004a}. This may be the result of small differences in the impact scenarios or simulation details, but the similarity between our two different clump iron core mass fractions is striking, particularly given that the iron core of the Moon itself has been inferred to be 1--2\% of the Moon's total mass \citep{Viswanathan+2019}.

The radial variation of the provenance of the rock within the clump has implications for predictions of the isotopic composition of lunar samples~\citep[c.f.][]{Cano+2020}. ~\autoref{fig:composition_clumps} shows how the mass fraction of proto-Earth increases linearly towards the surface of the clump. The fractional distance from centre to surface is computed using an ellipsoidal surface defined by the inertia tensor to account for the tidal distortion of the $l_{\rm Th}=0$ proto-Moon. Roughly equal amounts of Theia and proto-Earth are found at the surface of the clump; quite different from the overall 70/30 split. The results for the clumps in the $l_{\rm Th}=0$ and $\tfrac{1}{4}$ simulations continue to be very similar.

Outside the Roche radius, in addition to any large clumps there is a diffuse debris disc produced in each of our five simulations. The composition of these discs is shown in~\autoref{fig:composition}, split by material and provenance. There is more mass in the debris disc than in resolved orbiting clumps in the two counter-rotating scenarios, with a more massive disc being formed by the impact with the less rapidly counter-rotating Theia. For the discs formed in the $l_{\rm Th}=0$ and $\tfrac{1}{4}$ simulations, the overall bound mass exterior to the Roche radius grows with increasing $l_{\rm Th}$. However, the balance shifts from having a more massive disc to a more massive clump such that the disc mass decreases with increasing $l_{\rm Th}$ in this regime. As the clump, or at least its centre, is rich in Theia, the debris disc becomes less dominated by Theia, with almost equal amounts of target and impactor mantle when $l_{\rm Th}=\tfrac{1}{4}$. Thus, the disc material that may subsequently be accreted by the orbiting clump has a composition that is very similar to that already present at the surface of the clump.

When performing the same simulations with $10^6$ particles we find the outcomes to be significantly different to those presented here for the simulations containing $10^7$ particles. For instance, the simulation with $l_{\rm Th}=0$ does not yield an orbiting clump; instead it creates a merger similar to the $l_{\rm Th}=-\tfrac{1}{4}$ and $l_{\rm Th}=-\tfrac{1}{2}$ simulations. Also, while the composition of the mass in bound orbits exterior to the Roche radius is robust to this change in numerical resolution, the amount of this material is $\sim 45\%$ larger in the higher resolution case.

\section{Conclusions}
\label{conclusions}

We have presented a method to compute the hydrostatic equilibrium state of a uniformly rotating compressible fluid object, described as a set of concentric oblate spheroids, each with constant density (\cref{sec:hydro}). We then described an adaptation of the SEAGen algorithm of \citet{Kegerreis+2019} that places particles to match precisely this equilibrium configuration (\cref{sec:picleplacement}). The combination of these two tasks is performed by our new, open-source code WoMa, implemented in python and publicly available at \url{https://github.com/srbonilla/WoMa} and the python module \texttt{woma} can be installed directly with \href{https://pypi.org/project/woma/}{pip}. We tested its capabilities using simulations containing up to just over $10^9$ SPH particles that were evolved with the SWIFT code. Relative to previous studies that make particle-based models of rotating planets by incrementally adding rotation between repeated relaxation simulations, our method has the advantages of being fast and allowing precise control over the structure of the rotating planet to be simulated.

We used this new technique to study the effect of different rotation rates of Theia in $10^7$-particle simulations of a canonical Moon-forming impact. Counter-rotating Theia's produced quick mergers, whereas a rapidly corotating Theia led to a hit-and-run collision with numerous unbound clumps escaping from the Earth. In the zero spin and slowly corotating Theia cases, a large clump was left orbiting the Earth after 100 hours. The mass and composition of the resulting debris disk also varies systematically with the initial spin of Theia. Our findings confirm previous results, using lower resolution studies, that the outcomes of planetary giant impacts can depend strongly on the initial spins of the colliding bodies. Therefore, N-body simulations that aim to trace the formation of terrestrial planetary systems using models for the aftermath of giant impacts should track the spin of the forming planets.

The simulation with Theia not spinning initially yields an orbiting proto-Moon with a periapse at $4.5~R_\oplus$, well outside the Roche radius. It has a mass of $0.01~M_\oplus\,\simeq~0.81~M_{\rm \leftmoon}$, of which $\sim$1\% is an iron core, and while its overall fraction of proto-Earth material is only 30\%, a radial gradient in material provenance means that $\sim$50\% of the surface material originates in the proto-Earth. This fraction is similar to that in the Roche-exterior debris disc surrounding the Earth. Compared with previous studies of the canonical impact, which typically found a proto-Earth fraction of only $\sim$30\% in the potential Moon-forming material, our value is higher because some of Theia is already hidden deep within the proto-Moon. These shared characteristics suggest that this proto-Moon might be a plausible route for forming the Moon.

We also find that the results from our Moon-forming giant impact simulations can be sensitive to numerical resolution, with the collision outcome changing in one out of the five scenarios when increasing the particle number from $10^6$ to $10^7$. However, while the results presented here have not demonstrated numerical convergence, they do use more particles than has been typical in studies of the Moon-forming impact and we are running higher resolution simulations to test for numerical convergence.

There are also untested uncertainties associated with: the formulation of SPH being used, where artificial clumping could arise as a consequence of discontinuities in the density field and the level of material mixing will also be affected; and the choice of EoS, which will influence the detailed structure of the debris disc and the depletion of volatile elements accreting onto the proto-Moon. In future work, we will address these uncertainties as well as investigating a larger volume of parameter space for the impact scenario to determine how common these orbiting proto-Moons are.

\section*{Acknowledgements}

SRB is supported by a PhD Studentship from the Durham Centre for Doctoral Training in Data Intensive Science, funded by the UK Science and Technology Facilities Council (STFC, ST/P006744/1) and Durham University. VRE and RJM acknowledge support from the STFC grant ST/P000541/1. JAK is supported by STFC grants ST/N001494/1 and ST/T002565/1. 

This work used the DiRAC@Durham facility managed by the Institute for Computational Cosmology on behalf of the STFC DiRAC HPC Facility (www.dirac.ac.uk). This equipment was funded by BIS National E-infrastructure capital grant ST/K00042X/1, STFC capital grants ST/H008519/1 and ST/K00087X/1, STFC DiRAC Operations grant ST/K003267/1 and Durham University.

The research in this paper made use of the SWIFT open-source simulation code \citep[\href{http://www.swiftsim.com}{www.swiftsim.com},][]{Schaller+2016} version 0.8.1.





\bibliographystyle{mnras}
\bibliography{gihr}




\appendix




\bsp	
\label{lastpage}
\end{document}